\documentclass[10pt,twocolumn,letterpaper]{article}

\usepackage[pagenumbers]{cvpr} % To force page numbers, e.g. for an arXiv version

\newcommand{\CUT}[1]{}

\usepackage{color,xcolor}
\usepackage{multirow} 
\usepackage{booktabs} 
\usepackage{colortbl}
\definecolor{mygray}{gray}{.9}

\usepackage{pifont}
\usepackage{makecell}
\usepackage{graphicx}
\usepackage{amsmath}
\usepackage{booktabs}
\usepackage{wrapfig}
\usepackage{multirow}  
\usepackage{bm} \bm{}
\usepackage[linesnumbered,ruled,vlined]{algorithm2e}

\def\z{{\bm z}}

\definecolor{cvprblue}{rgb}{0.21,0.49,0.74}
\usepackage[pagebackref,breaklinks,colorlinks,allcolors=cvprblue]{hyperref}

\def\paperID{*****} % *** Enter the Paper ID here
\def\confName{CVPR}
\def\confYear{2026}

\title{OmniCustom: Sync Audio-Video Customization Via Joint Audio-Video Generation Model}

\author{  \\
[-5pt]
Maomao Li${^{1}}$,\quad Zhen Li${^{2}}$,\quad Kaipeng Zhang${^{2}}$, \quad Guosheng Yin${^{1}}$, \quad Zhifeng Li${^{3}}$ \quad Dong Xu${^{1}}$\\
[5pt]
${^1}$The University of Hong Kong \qquad
${^2}$Shanda AI Research Tokyo \qquad \\
${^3}$XIntelligence Technology Co., Limited  \\
{\tt\small limaomao07@connect.hku.hk \qquad \{li.zhen1,\ kp\_zhang\}@foxmail.com} \\  {\tt\small zhifeng0.li@gmail.com} \qquad
{\tt\small \{gyin,\ dongxu\}@hku.hk}
}

\begin{document}

\twocolumn[{%
\renewcommand\twocolumn[1][]{#1}%
\maketitle
\begin{center}
    \centering
    \captionsetup{type=figure}
    \includegraphics[width=\textwidth]{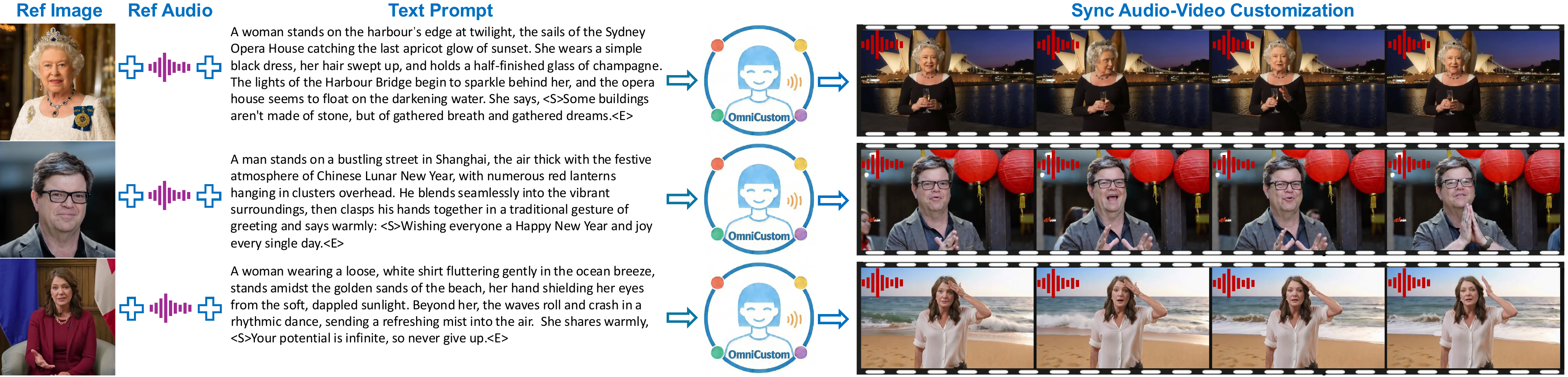}
   \vspace{-2em}
       \caption{We propose OmniCustom, a novel framework to deal with sync audio-video customization. Given a reference image $I^{r}$ and a reference audio $A^{r}$, OmniCustom synchronously generates a video that preserves the identity from $I^{r}$ and an audio that mimics the timbre of $A^{r}$. The speech content can be specified by a text prompt, where we use $<$S$>$ and $<$E$>$ to mark the start and end of a speech.
       }  
    \label{fig:teaser}
\end{center}%
}]
\begin{abstract}
Existing mainstream video customization methods focus on generating identity-consistent videos based on given reference images and textual prompts. Benefiting from the rapid advancement of joint audio-video generation, this paper proposes a more compelling new task: \textit{sync audio-video customization}, which aims to synchronously customize both video identity and audio timbre. Specifically, given a reference image $I^{r}$ and a reference audio $A^{r}$, this novel task requires generating videos that maintain the identity of the reference image while imitating the timbre of the reference audio, with spoken content freely specifiable through user-provided textual prompts. To this end, we propose OmniCustom, a powerful DiT-based audio-video customization framework that can synthesize a video following reference image identity, audio timbre, and text prompts all at once in a zero-shot manner. Our framework is built on three key contributions. First, identity and audio timbre control are achieved through separate reference identity and audio LoRA modules that operate through self-attention layers within the base audio-video generation model. Second, we introduce a contrastive learning objective alongside the standard flow matching objective. It uses predicted flows conditioned on reference inputs as positive examples and those without reference conditions as negative examples, thereby enhancing the model’s ability to preserve identity and timbre. Third, we train OmniCustom on our constructed large-scale, high-quality audio-visual human dataset. Extensive experiments demonstrate that OmniCustom outperforms existing methods in generating audio-video content with consistent identity and timbre fidelity. Project page: \url{https://omnicustom-project.github.io/page/}.
\end{abstract}

% we introduce two lightweight condition injection LoRA Module on a sync audio-video generation model.

% The task is extremely challenging, but it provides a more complete customization capability over the generated audio-video samples regarding video identity and audio timbre. 

\begin{table*}[t]
\footnotesize
    \centering
    \begin{tabular}{l|c|c|c|c}
    \toprule  
      % Settings   &\makecell[c]{Identity\\preservation} &\makecell[c]{Audio\\containing} &\makecell[c]{Audio\\customization} &\makecell[c]{Background \\Sounds} \\
      Settings   & {Identity preservation} &{Audio containing} &{Audio customization} &{Background Sounds} \\
    \hline
 {Typical video customization} & \ding{51}  &\ding{55}  &\ding{55} &\ding{55}\\
\hline
{Audio-driven video customization}  &\ding{51}  &\ding{51} &\ding{55} &\ding{55}\\
\hline
 \rowcolor{mygray} {Sync audio-video customization} &\ding{51}  &\ding{51}  &\ding{51} &\ding{51}\\
     \bottomrule
    \end{tabular}   
    \vspace{-0.8em}
    \caption{Categories and characteristics of video customization methods. We first propose \textit{sync audio-video customization}, which jointly produces identity-consistent videos and timbre-cloned audio tracks. Current audio-driven video customization approaches output videos with paired audio, but must regenerate the driving audio to adjust spoken content. Additionally, our method features a unique strength: the generation of contextually relevant background sound effects.}
\label{tab:setting}
     \vspace{-1.2em}
\end{table*}

% We propose SyncCustom, a novel framework to deal with sync audio-video customization. Specifically, given a reference image $I_{ref}$ and a reference audio $A_{ref}$, our goal is to synchronously generate a identity consistent video following the ID from $I_{ref}$ and a speech audio following user-given speech context with cloned timbre from $A_{ref}$. 
\section{Introduction}
\label{sec:intro}
Deep generative models~\cite{goodfellow2020generative,kingma2018glow,ho2020denoising,liu2022flow,lipman2022flow} have demonstrated a remarkable capacity for producing high-quality samples across various data modalities. Leveraging large-scale training data and powerful architectures, text-to-video (T2V)~\cite{ge2023preserve,gu2023reuse,an2023latent,blattmann2023align,guo2023animatediff,yang2024cogvideox,kong2024hunyuanvideo,wan2025wan} have enabled the synthesis of vivid visual content from textual descriptions.  Building upon these advances, video customization ~\cite{jiang2024videobooth,he2024id,wang2024customvideo,wei2024dreamvideo,liu2025phantom,fei2025skyreels,jiang2025vace} aim at synthesizing identity-preserving videos, which garners increasing attention due to their broad potential for applications ranging from film and advertising to gaming.

Existing video customization techniques can be classified into two categories: tuning‑based and tuning‑free. Tuning‑based methods~\cite{ma2024magic,wang2024customvideo,wei2024dreamvideo,wu2024motionbooth} require fine‑tuning the pre‑trained model for each new identity during inference. For instance, DreamVideo~\cite{wei2024dreamvideo} simultaneously customizes identity and motion through separate identity and motion adapters. While these approaches have shown promising results, their reliance on per‑identity tuning during inference limits practical scalability. In contrast, tuning‑free methods~\cite{he2024id,huang2025conceptmaster,yuan2025identity,jiang2025vace} can introduce new identities without additional test-time training. For example, ID‑Animator~\cite{he2024id} employs learnable facial latent queries to extract identity embeddings, enabling zero‑shot video generation. ConsisID~\cite{yuan2025identity} further leverages control signals derived from frequency decomposition, where low‑frequency features guide pixel‑level prediction and high‑frequency cues help preserve fine facial details.
Moreover, several recent studies~\cite{liu2025phantom,fei2025skyreels,jiang2025vace,huang2025conceptmaster} have also explored multi‑object customization in video synthesis.

Although fruitful results have been achieved, typical video customization only produce silent outputs. Recently, another line of methods~\cite{yi2025magicinfinite,hu2025hunyuancustom,chen2025humo,wang2025interacthuman} incorporate audio as an additional modality, enabling audio-driven video customization. However, it is cumbersome for these two-stage customization methods to modify spoken content while retaining a given timbre, as they require regenerating a driven audio via Text-to-Speech (TTS) techniques~\cite{wang2023neural,zhang2023speak,le2023voicebox}. Additionally, adopting TTS outputs is incompatible with scenarios containing background sound effects (\eg, beach), since background sounds cannot be synthesized by TTS models. Given the above considerations, this paper introduces a novel task termed \textit{sync audio-video customization}, motivated by a simple observation: humans naturally associate personal identity with voice timbre.
Specifically, as illustrated in Fig.~\ref{fig:teaser}, given a reference image $I^{r}$, a reference audio $A^{r}$, and a textual prompt, our goal is to synchronously generate a video that preserves the visual identity of $I^{r}$ and a corresponding audio that mimics the timbre of $A^{r}$, where spoken content can be freely given via textual prompt. Tab.~\ref{tab:setting} summarizes the functional differences between existing customization techniques and our proposed \textit{sync audio-video customization}. Although this new task poses great challenges,
recent advances in joint audio-video generation~\cite{veo3,sora2,wang2025universe,low2025ovi}, particularly the open-source OVI~\cite{low2025ovi}, have laid the groundwork for its feasibility and potential background sounds.

This paper presents OmniCustom, an efficient reference-guided Diffusion Transformer (DiT) framework~\cite{peebles2023scalable} that achieves \textit{sync audio-video customization} through several key innovations.
First, our method introduces separate reference image and audio branches into the original video and audio streams within the fusion block of the OVI architecture~\cite{low2025ovi}. To maintain efficiency, we incorporate two independent LoRAs~\cite{hu2022lora} into the QKV projections of the reference tokens, thereby avoiding massive computational overhead. Second, our approach is trained with a compound objective that pairs the standard flow matching loss with an auxiliary contrastive learning objective, which maximizes the dissimilarity between predicted flows from samples with reference conditions and those without, thus enhancing identity and timbre preservation. 
Leveraging these techniques, we train OmniCustom on OmniCustom-1M, which is a large-scale, high-quality audio-visual human dataset we constructed, comprising one million single-human portrait videos. With its rich annotations and standardized format, we anticipate that this dataset will serve as a foundational resource for future work in \textit{sync audio-video customization} and related applications. In summary, our main contributions are as follows:
\begin{itemize}
\item 
We introduce OmniCustom, a tuning-free sync audio-video customization model that can generate a personalized video preserving the identity from reference image $I^{r}$ and an audio mimicking the timbre from reference audio $A^{r}$.
\item
OmniCustom incorporates reference image and audio branches into the original video and audio streams in OVI, respectively. To improve the fidelity of video identity and audio timbre, we design a contrastive learning objective that uses samples with reference conditions as positive examples and those without as negative examples.
\item
We construct a large-scale audio-visual human dataset comprising 1 million examples to train the proposed OmniCustom. Extensive qualitative and quantitative experiments demonstrate that our method yields high-quality, identity-consistent videos with timbre-cloned audio tracks.
\end{itemize}

\section{Related Works}

\noindent{\textbf{Joint Audio-Video Generation.}}
Joint audio-visual generation has witnessed rapid advancement in recent years. Specifically, MM-Diffusion~\cite{ruan2023mm} is the first attempt, which consists of a sequential multi-modal U-Net, where two subnets for audio and video learn to gradually generate aligned audio-video pairs from Gaussian noises. However, the model is unconditional and trained on limited data scope, \eg, landscapes~\cite{landscapes} and dancing~\cite{li2021ai}, leading to insufficient generalization ability. Afterwards, Seeing-and-Hearing~\cite{xing2024seeing} realizes text-guided joint video-audio generation by applying ImageBind~\cite{imagebind} as an aligner in the diffusion latent space of different modalities. Nevertheless, it sometimes results in low-quality and temporally misaligned output. Recently, Veo 3~\cite{veo3} and Sora 2~\cite{sora2} demonstrate new milestone performance of sync audio-video generation. As a representative of open source models, OVI~\cite{low2025ovi} trains an audio backbone from scratch using MMAudio~\cite{cheng2025mmaudio}, and then achieves audio-video fusion via paired cross-attention layers. 

% UniVerse-1~\cite{wang2025universe} and Ovi~\cite{low2025ovi} propose open-source models, where the former employs a music generation model ACE-Step~\cite{gong2025ace} as the audio backbone while the latter trains a aduio backbone from scratch using MMAudio~\cite{cheng2025mmaudio}.

\noindent{\textbf{Conditional Audio Generation.}} Generative audio modeling is a well-established field, which usually contains speech~\cite{shen2018natural,li2019neural,ren2020fastspeech,kim2020glow,popov2021grad} and environmental sounds~\cite{yang2023diffsound,huang2023make,liu2024audioldm}. Specifically, Text-to-Speech (TTS)~\cite{wang2023neural,zhang2023speak,le2023voicebox} aims to synthesize speech for any given text and mimic the speaker of audio prompt. Here, diffusion TTS methods~\cite{shen2023naturalspeech,ju2024naturalspeech,chen2025f5,du2024cosyvoice} can conduct parallel processing for fast inference. Further, another group of methods~\cite{peng2024voicecraft,han2024vall,du2025vall,liao2024fish} use an autoregressive (AR) architecture, which consecutively predicts next tokens for zero-shot TTS capability.

\noindent{\textbf{Identity-preserving Video Customization.}}
In recent years, the architecture of video customization has converted from UNet to Transformer-based DiT~\cite{peebles2023scalable}. As a representative method during the UNet-based era, MagicMe~\cite{ma2024magic} employs separate training for each personalized ID, making it difficult to achieve zero-shot capabilities. As a zero-shot human-video generation approach, ID-Animator~\cite{he2024id} can perform personalized generation given a single reference facial image without further training. The DiT architecture has driven the field toward token-based approaches~\cite{yuan2025identity,jiang2025vace,liu2025phantom}. For example, VACE~\cite{jiang2025vace} proposes an all-in-one model for video creation that uses a pluggable Context Adapter to inject concepts from different tasks into the model.
Phantom~\cite{liu2025phantom} enhances the joint text-image injection mechanism and trains it to capture cross-modal correspondences using text-image-video triplet data. In addition, there are also a number of methods that focus on 
multi-object customization~\cite{liu2025phantom,fei2025skyreels,jiang2025vace,huang2025conceptmaster}, audio-driven video customization~\cite{yi2025magicinfinite,hu2025hunyuancustom,chen2025humo,wang2025interacthuman} and image customization~\cite{yue2024addme,ye2023ip,guo2024pulid,qian2025composeme,yuan2023inserting}.

% In addition, several methods focus on audio-driven video customization~\cite{yi2025magicinfinite,hu2025hunyuancustom,chen2025humo,wang2025interacthuman} and customized image generation~\cite{yue2024addme,ye2023ip,guo2024pulid,qian2025composeme}. 

%加数字人

Unlike existing works, this paper proposes a new task: \textit{sync audio-video customization}, which enables simultaneous customization of both visual identity and audio timbre. Compared with audio-driven customization techniques, our task offers greater flexibility by allowing users to freely specify spoken content through textual prompts.

\begin{figure*}[t]
  \centering
  \includegraphics[width=1\linewidth]{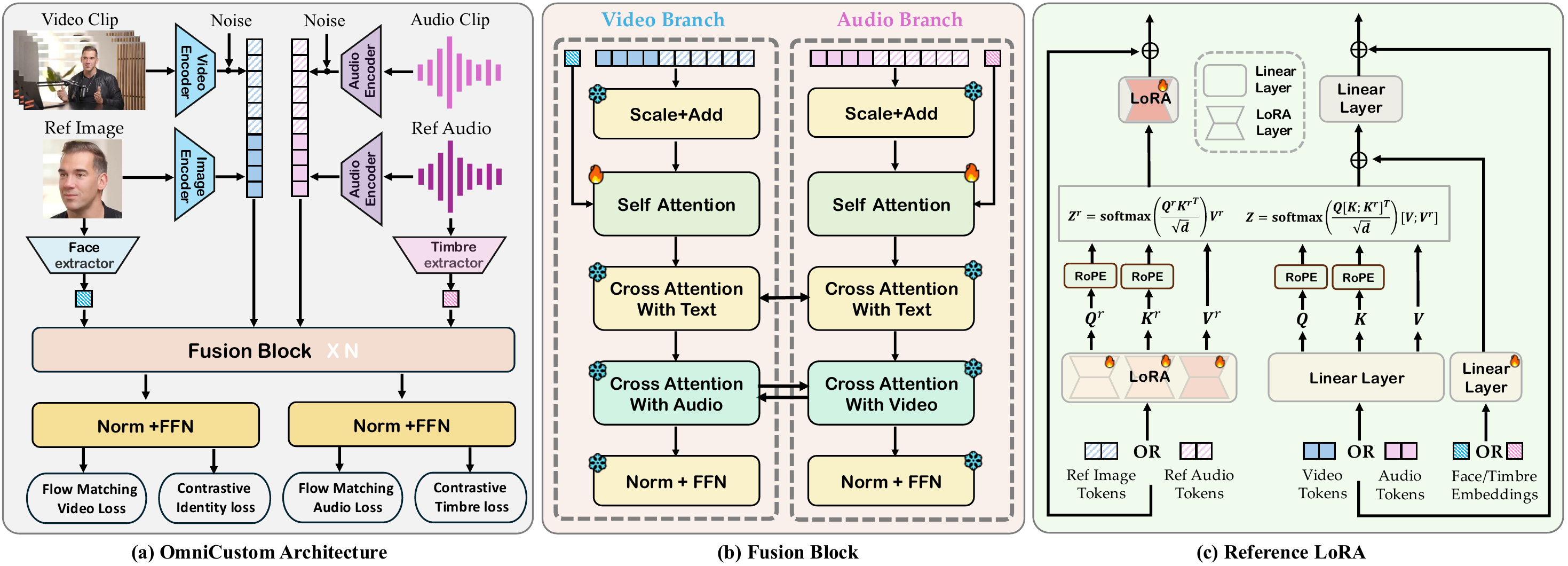}
        \vspace{-1.2em}
  \caption{(a) Overview of our OmniCustom architecture. We extend the joint audio-video generation model OVI~\cite{low2025ovi} by introducing reference image and audio branches alongside the original video and audio flows. The visual and audio VAE encoders project the reference image $I^{r}$ and audio $A^{r}$ into tokens, which are then concatenated with the noised video and audio latent tokens, respectively, before being processed by the fusion blocks. Here, the face embeddings~\cite{deng2019arcface} and timbre embeddings~\cite{ju2024naturalspeech} are also input in fusion blocks for further constraint. (b) Fusion Block. It is designed as a symmetric twin backbone with parallel audio and video branches. Our OmniCustom embraces identity and timbre information via finetuning self-attention layers in video and audio branches, respectively. (c) Reference LoRAs. We incorporate separate LoRA into the QKV projections of the reference identity and audio representations. Specifically, the reference identity LoRA is in the self-attention layers of video branch, whereas the reference audio LoRA is in those of audio branch.  
  }
  \label{fig:pipeline}
\end{figure*}

% (a) The architecture of the proposed OmniCustom. We introduce a reference image and a reference audio branch alongside the original video and audio branch in the base joint audio-video generation model OVI. Visual and audio VAE encoder maps reference image $I^{r}$ and  $A^{r}$ into tokens, which are concatenated with the noised video and audio latent tokens respectively, and then passed through the fusion blocks. (b) The fusion blocks is designed as symmetric twin backbone with parallel audio and video branches. Our OmniCustom embraces identity and timbre information via finetuning self-attention layers in video and audio branch, respectively. (c) We incorporate separate LoRAs into the QKV projection of reference identity and reference audio representations. Here, the reference identity LoRA is within self-attention layers in the video branch, while the reference audio LoRA is within that in the audio branch.
\section{Preliminary} 
% \subsection{Synchronous Audio-Video Generation}
\noindent{\textbf{Synchronous Audio-Video Generation.}}
OVI~\cite{low2025ovi} is a text‑guided synchronous audio‑video generation model that achieves similar functionality with Veo 3~\cite{veo3} and Sora 2~\cite{sora2}. It adopts a twin backbone with parallel audio and video DiT branches. The video branch is initialized from Wan2.2 5B~\cite{wan2025wan}, while the structurally identical audio branch is trained from scratch using the pre‑trained 1D VAE from MMAudio~\cite{cheng2025mmaudio}. As illustrated in Fig.~\ref{fig:pipeline} (b), OVI’s fusion block employs not only standard text cross‑attention but also paired cross‑attention layers for audio‑video fusion, allowing the audio stream to attend to the video stream and vice versa. 
This bidirectional mechanism continuously propagates synchronization signals throughout the entire network.

% \subsection{Flow-matching Objective}
\noindent{\textbf{Flow-matching Objective.}}
OVI~\cite{low2025ovi} and Wan2.2~\cite{wan2025wan} applies the flow matching loss~\cite{liu2022flow,lipman2022flow}, which learns a straight flow trajectory between data and noise distributions. With a Gaussian latent $\z_1 \in \mathcal{N}(0,I)$, the forward process linearly corrupts the clean latent $\z_0$ as:
\begin{equation}
\begin{aligned}
\z_t = (1-t)\z_0 + t \z_1,
\end{aligned}
\label{forward}
\end{equation}
where $t$ is sampled from a uniform distribution. Then, the backward process learns a velocity field $v_{\theta}(\cdot)$ that maps samples from the Gaussian distribution to the data distribution. This is formalized as a least squares regression problem, where $v_{\theta}$ is optimized to approximate $\z_1 - \z_0$:
\begin{equation}
\begin{aligned}
\min_{\theta}\int_{0}^{1}\mathbb{E}[\parallel v_{\theta}(\z_t,t) - (\z_1-\z_0)\parallel^2]dt.
\end{aligned}
\label{rf}
\end{equation}
After training the velocity field $v_{\theta}$, the model can reconstruct a clean latent ${\z}_0$ from pure noise $\z_1$, where the sampling procedure is with a discrete sequence of $N$ time steps $t_i$:
\begin{equation}
\begin{aligned}
\z_{t_{i-1}} = \z_{t_i} + (t_{i-1}-t_i)v_{\theta}(\z_{t_i},t_i).
\end{aligned}
\label{rf_sample}
\end{equation}

% \subsection{Subject Customization}

% \input{fig_tex/pipeline}
\section{Methods}
\label{sec:method}

\subsection{Problem Formulation}
Given a reference image $I^{r}$ and a reference audio clip $A^{r}$, we formally define a novel task termed \textit{sync audio-video customization}. Its core objective is to simultaneously generate two aligned outputs: a video that preserves the identity information derived from $I^{r}$, and an audio that mimics the timbre characteristics of $A^{r}$, where speech content can be freely specified by the user-given text prompt.

Our proposed task differs fundamentally from existing audio-driven customization \cite{hu2025hunyuancustom,chen2025humo}, despite the fact that both paradigms leverage a personalized image and audio, and a text prompt as input. Specifically, they synthesize audio-driven personalized videos, where the speech content and timbre are inherently predetermined by the input audio track. Here, 
to maintain given timbre and specify a new speech context, users need to apply Text-to-Speech (TTS)~\cite{wang2023neural,zhang2023speak,le2023voicebox} techniques to generate a new driven audio in advance, which is quite troublesome. In contrast, our framework enables simultaneous customization of both video identity and vocal timbre \textbf{in a single pass}, thereby granting significantly greater flexibility in the customization pipeline. Furthermore, benefiting from the base model of joint audio-video generation, our method is also capable of producing contextually relevant background sound effects (\eg, ocean wave sound).

% The rest of this section is organsized as follows.  In Sec.~\ref{4.2}, we first present the model design of our SyncCustom. Then we provide the explanation of our training strategy and data augmentation techniques in Sec.~\ref{4.3}.
% SyncCustom is a multi-modal customized generation model built upon a sync audio-video generation model OVI~\cite{low2025ovi}. It can generate a personalized video from reference image, audio, and text prompt, while maintaining consistency of reference identity and audio timbre. 

\subsection{Model Designs}
\label{4.2}
The overall architecture of our OmniCustom is depicted in Fig.~\ref{fig:pipeline} (a), where we adopt OVI~\cite{low2025ovi} as the sync audio-video generation backbone. To enable simultaneous conditional information integration derived from the reference image $I^{r}$ and reference audio $A^{r}$, we introduce two dedicated reference branches that operate in parallel with the original video and audio branches within the fusion module of the OVI architecture. Rather than introducing extra auxiliary modules, we integrate LoRA~\cite{hu2022lora} into the QKV projection layers of the reference image and audio tokens. This design preserves the inherent structure of the OVI model while avoiding excessive trainable parameters and heavy computational overhead. Furthermore, to achieve stronger reference constraints, we extract facial and timbre embeddings~\cite{deng2019arcface,ju2024naturalspeech} during training. Additionally, we design two contrastive learning objectives that maximize the dissimilarity between the predicted flows generated with reference conditions and those without.

% The overall framework of the proposed OmniCustom is illustrated in Fig.~\ref{fig:pipeline}, where we use OVI~\cite{low2025ovi} as the base sync audio-video generation model. To achieve conditional integration from reference image $I^{r}$ and reference audio $A^{r}$ simultaneously, this paper introduces two reference branch alongside the original video and audio branch in the fusion block in OVI architecture. Instead of applying additional modules, we incorporate LoRA~\cite{hu2022lora} into the QKV projection of reference image and audio tokens, which maintains OVI architecture while avoiding massive training parameters and computational overhead. Besides, our approach adds a adaptive contrastive objective that maximizes dissimilarities between predicted flows from the one without reference samples.  

\subsubsection{OmniCustom Architecture}
\label{sec:4.2.1}

As shown in Fig.~\ref{fig:pipeline} (a), we first encode the reference image $I^{r}$ into the latent space using an encoder of pre-trained Variational Autoencoder (VAE)~\cite{kingma2013auto}. 
Subsequently, these image latents undergo the same patchification and encoding procedures that are applied to the video latents. For the reference audio $A^{r}$, we convert the raw audio signal via Short-Time Fourier Transform (STFT) and extract mel-spectrograms~\cite{Stevens1937ASF}, which are then encoded into latents by leveraging the pre-trained 1D VAE from MMAudio~\cite{cheng2025mmaudio}. Note that we treat the reference image and audio as static conditional inputs to maintain a time-invariant nature, where we assign time step of $0 $ to these reference conditions.

The reference image and audio tokens are then concatenated with the original video and audio latent tokens, respectively. These concatenated token sequences are subsequently fed into the fusion module of the OVI model, as illustrated in Fig.~\ref{fig:pipeline} (b). Specifically, this fusion module adopts a twin-backbone architecture with parallel audio and video branches, where each layer is fully symmetric and configured with an identical number of transformer blocks. Paired cross-attention layers are integrated to facilitate mutual information exchange between audio and video branches, where both modalities can attend to each other effectively.
In this paper, we only fine-tune the unimodal self-attention modules in fusion blocks, which allows reference representation injection while preserving audio-video alignment.
The reference LoRA is detailed in Sec.~\ref{sec:4.2.2}. Finally, the generated video latents are decoded into the pixel space, whereas the corresponding audio latents are first decoded into waveforms via the 1D VAE from MMAudio~\cite{cheng2025mmaudio}, and then further vocoded into high-fidelity audio waveforms using a pre-trained vocoder~\cite{lee2022bigvgan}.

\subsubsection{Reference LoRA}
\label{sec:4.2.2}
We start by projecting the original input video and audio features ($X$) into query ($Q$), key ($K$), and value ($V$) features. Specifically, these features are derived through the self-attention mechanism of the Transformer, following the standard QKV projection as formulated below:
\begin{equation}
\begin{aligned}
Q, K, V = W_QX,~ W_KX, ~W_VX,
\end{aligned}
\label{eqn:QKV}
\end{equation}
where $W_Q$, $W_K$, and $W_V$ are projection matrices. Positional information is injected into $Q$ and $K$ using RoPE~\cite{su2024roformer} before self-attention.

To efficiently integrate reference conditions while preserving the generalization capability of the pre-trained model, we extend the OVI architecture by introducing two dedicated reference branches. As illustrated in Fig.~\ref{fig:pipeline} (c), we deploy two separate LoRAs~\cite{hu2022lora} to handle two distinct modalities, respectively. Specifically, the audio and video tokens are processed independently, where reference-image tokens and original video tokens are fed into the self-attention layers of the video branch, whereas reference-audio tokens and original audio tokens are routed to those in the audio branch. The corresponding projections of reference features ($X^r$) are as follows:
\begin{equation}
\begin{aligned}
Q^r &= (W_Q + B_Q A_Q)X^r, \\
K^r &= (W_K + B_K A_K)X^r, \\
V^r &= (W_V + B_V A_V)X^r,
\end{aligned}
\label{eqn:lora}
\end{equation}
where $r \in \{I^{r}, A^{r}\}$. $A_i$ and $B_i$ ($i \in \{Q, K, V\}$) are low-rank matrices in $\mathbb{R}^{n \times d}$ and $\mathbb{R}^{d \times n}$ separately. Here, $n \ll d$, parameterizing the LoRA transformation. RoPE~\cite{su2024roformer} is also applied to $Q^r$ and $K^r$.
Consequently, the resulting reference-image or reference-audio features (\ie, $Z^r$) and the resulting features of video or audio branch (\ie, $Z$) are derived as:
\begin{equation}
\begin{aligned}
Z^r &= {\rm softmax} ({Q^r{K^r}^{\top}}/{\sqrt{d}}) V^r, \\
Z &= {\rm softmax} ({Q[K;K^{r}]^{\top}}/{\sqrt{d}}) [V; V^r],
\end{aligned}
\label{eqn:z}
\end{equation}
where $[\cdot~;~\cdot]$ denotes concatenation along the sequential dimension. Furthermore, to reinforce identity and timbre signals, we use InsightFace~\cite{deng2019arcface} and Naturalspeech 3~\cite{ju2024naturalspeech} to extract facial embeddings ($\mathbf{E}_f \in \mathbb{R}^{1\times 512}$) and timbre embeddings ($\mathbf{E}_t \in \mathbb{R}^{1\times 256}$). These reference embeddings are processed independently via several trainable linear layers. Then, the projected identity features are added to the self-attention output features $Z$ in the video branch, while the projected timbre features are added to $Z$ in the audio branch. The process can be expressed as:
\begin{equation}
\begin{aligned}
Z_o=Z+\varepsilon_{\phi}(\mathbf{E}), 
\end{aligned}
\label{add}
\end{equation}
where $\varepsilon_{\phi}(.)$ is trainable learner layers, $\mathbf{E}$ indicates corresponding reference embeddings.
Note that the facial embeddings and timbre embeddings are a single feature, which serves as a global condition for each branch. Next, we apply LoRA to the output linear layer to further reduce parameters involved in fine-tuning.

% Besides, to further enhance personalized identity, we extract 512-d face embeddings~\cite{ren2023facial} and send them into a linear layer, which are then added with the resulting features $Z$ of video. Similarly, we use a timbre extractor~\cite{ju2024naturalspeech} for 256-d timbre features, and put them through a linear layer, which are then added with the resulting features $Z$ of audio.

% \input{fig_tex/cl_loss}

\subsubsection{Contrastive Learning Objective}
\label{sec:4.2.3}
In OVI, the flow-matching objective is applied to the audio and video modalities separately. Although no explicit synchronization loss is incorporated, the symmetric backbone facilitates audio–visual correspondences. In our OmniCustom framework, the flow-matching objective for the video branch can be formulated with reference image conditions:
\begin{equation}
\begin{aligned}
\mathcal{L}_{FM}^V=\mathbb{E}[\|v_{\theta}(Z_{t_i},I^r,C,t_i)-(Z_1-Z_0)\|^2],
\end{aligned}
\label{eqn:fm}
\end{equation}
where $C$ refers to text conditions. Similarly, the flow-matching objective for the audio branch, $\mathcal{L}_{FM}^A$, can be obtained using $A^r$.

\noindent{\textbf{The contrastive regularization.}}
To further enhance the identity and timbre preservation capabilities during training, we additionally introduce a contrastive learning objective as regularization. Contrastive learning is initially proposed for face recognition tasks~\cite{schroff2015facenet}, which imposes a discriminative margin between positive and negative face sample pairs. Recently, Contrastive Flow Matching~\cite{stoica2025contrastive} improves conditional separation performance, which explicitly enforces the uniqueness constraint across all conditional flows. However, to the best of our knowledge, contrastive learning
has not been explored in the context of  video customization under flow-matching objectives.

We introduce contrastive learning by using predicted flows with reference conditions as positive examples while those without reference conditions are negative examples. 
Specifically, the proposed contrastive identity objective $\mathcal{L}_{CL}^{I}$ tries to \textbf{push} the velocity field $v_{\theta}(Z_{t_i},I^r,C,t_i)$ conditioned on the reference image $I^r$ away from $v_{\theta}(Z_{t_i},\phi,C,t_i)$ conditioned only on text $C$ during training. This is achieved via the maximization of the following quantity: 
\begin{equation}
\begin{aligned}
\mathcal{L}_{CL}^{I}\!\!=\!\!-\mathbb{E}[\|v_{\theta}(Z_{t_i},I^r,C,t_i) \!\!-\!\! \texttt{StopGrad}(v_{\theta}(Z_{t_i},\phi,C,t_i)\|^2].
\end{aligned}
\label{eqn:cl}
\end{equation}
$\texttt{StopGrad}$ is the operation of stop gradient, which accelerates convergence and stabilize training~\cite{chen2020simple}.

Given the flow-matching objective for the video branch, $\mathcal{L}_{FM}^V$, attempts to \textbf{pull} the velocity field $v_{\theta}(Z_{t_i},I^r,C,t_i)$ conditioned on reference image $I^r$ and the optimized direction $(Z_1-Z_0)$ closer together, our 
contrastive identity objective $\mathcal{L}_{CL}^{I}$ can be regarded as regularization to it. Furthermore, contrastive timbre loss $\mathcal{L}_{CL}^{A}$ can be obtained in a similar fashion. Generally speaking, such design forces the model to learn the difference from non-reference injection, thus enhancing the ability of preserving given identity and timbre.
Here, positive examples are implied in the flow matching objectives $\mathcal{L}{FM}^V$ and $\mathcal{L}{FM}^A$. The no-reference injection for the negative examples involves two aspects. First, since the original audio-video branch attends to reference tokens via attention mechanism, we use masked attention in Eq.~(\ref{eqn:z}) to block reference information. Second, we set the facial and timbre embeddings to 0.

\noindent{\textbf{Putting them all together.}} 
The total loss of our OmniCustom is:
\begin{equation}
\begin{aligned}
\mathcal{L}_{\rm total} = \lambda_{V}\mathcal{L}_{FM}^{V} + \lambda_{A}\mathcal{L}_{FM}^{A} + \lambda_{I^r}\mathcal{L}_{CL}^{I} + \lambda_{A^r}\mathcal{L}_{CL}^{A},
\end{aligned}
\label{eqn:all_loss}
\end{equation}
where $\lambda_{V}$, $\lambda_{A}$, $\lambda_{I^r}$, and $\lambda_{A^r}$ are weight coefficients. The last two terms in Eq.~(\ref{eqn:all_loss}) can adjust the strength of the contrastive regularization.

\begin{figure*}[t]
  \centering
  \includegraphics[width=1\linewidth]{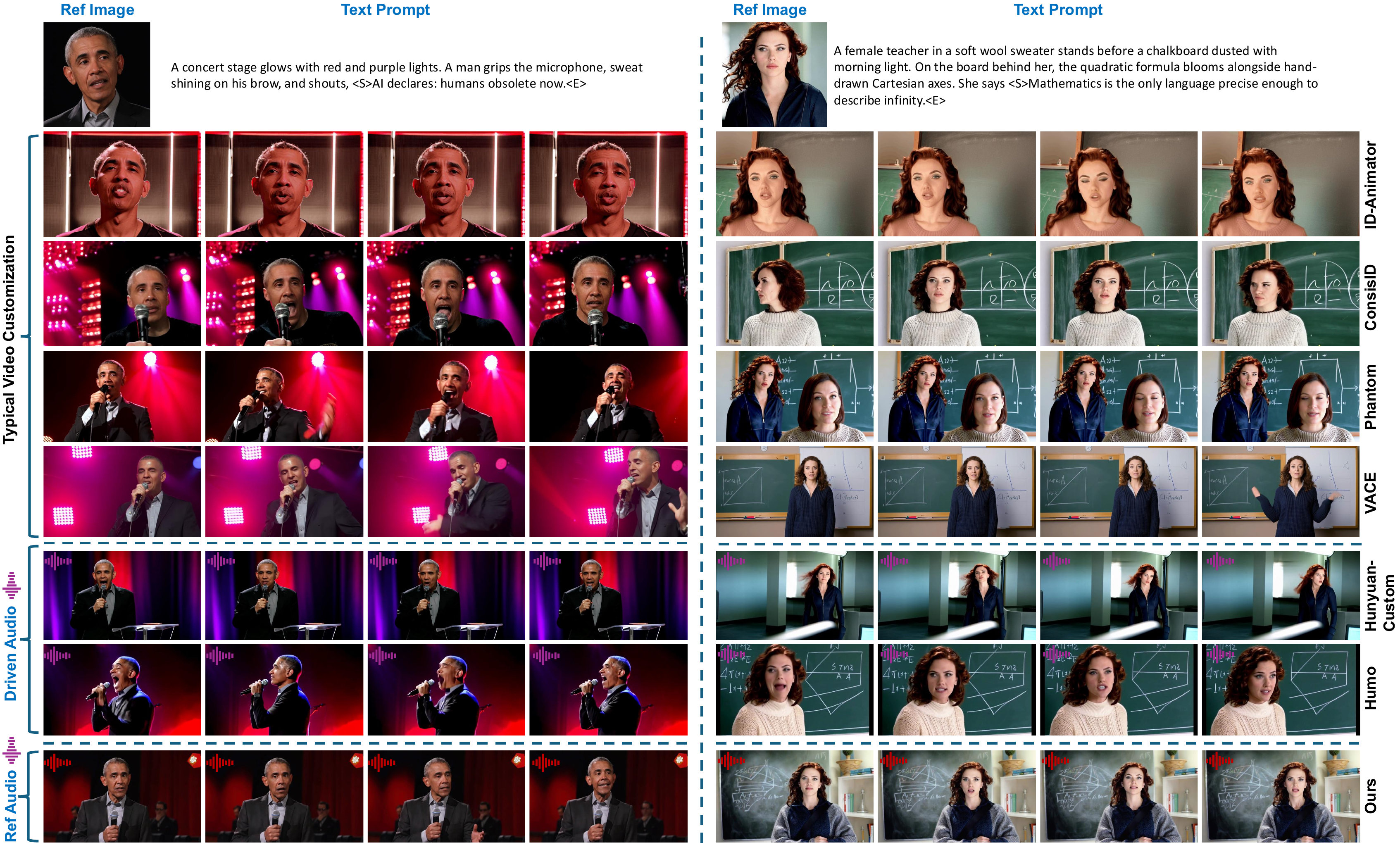}
        \vspace{-1em}
  \caption{Qualitative comparison with state-of-the-art video customization methods. The speech content of HunyuanCustom~\cite{hu2025hunyuancustom} and Humo~\cite{chen2025humo} is directly determined by their input audio. Our OmniCustom mimics the timbre of the reference audio and flexibly specifies the spoken content through textual prompts.
}
\label{fig:vis}
    % \vspace{-0.4em}
\end{figure*}

\section{Dataset: OmniCustom-1M} 
\label{5}
Driven by our proposed task, we create a large-scale, high-quality synchronous audio-video dataset OmniCustom-1M to fine-tune over joint audio–video model OVI~\cite{low2025ovi}.

\subsection{Dataset Sources} Our raw data is from SpeakerVid-5M~\cite{zhang2025speakervid}. Totaling over 8,000 hours, it contains more than 5.2 million video clips of human portraits talking. Each clip is accompanied by structured textual annotations, ASR transcriptions~\cite{radford2023robust}, and human bounding boxes, supporting multimodal learning objectives.

\subsection{Dataset Process} \noindent{\textbf{Filtering via Sync Detection.}} We use single-speaker audio-video clips in SpeakerVid-5M~\cite{zhang2025speakervid}. To seek better audio-video synchronization, we additionally filter videos by using offset and confidence score metrics of SyncNet~\cite{raina2022syncnet}: $|$ \textit{offset} $| \le 3$ and \textit{confidence} $> 1.5$. Additionally, to ensure the visual quality, we filter out videos with aesthetics scores~\cite{wu2023exploring} below 0.3.
This leads to a total number of roughly 1M single-person audio-video clips totaling 2,500 hours.

\noindent{\textbf{Audio Captioning.}} 
Although the SpeakerVid-5M~\cite{zhang2025speakervid} dataset includes video captions, it lacks audio captions. Following OVI~\cite{low2025ovi}, we construct audio captions for speech videos to emphasize the speaker's age, gender, accent, and vocal characteristics (\eg, pitch, prosody, emotion, and speaking rate).

\noindent{\textbf{Format Standardization.}}
We filter out videos shorter than 10 seconds. All videos are recorded in 480p at 24 FPS. We extract audio files from videos and unify them into 16kHZ.

\subsection{Division of Training Clips and Reference Clips}
Each training clip and its corresponding reference clip are sampled from the same video. Specifically, we extract the first 4 seconds as the reference audio from each segment in OmniCustom-1M. Subsequently, the last 5 seconds are designated as both the training audio and video clips. This setup ensures that each reference-training pair shares the same timbre but contains distinct speech content, thereby preventing the network from learning speech content instead of timbre. We use GLM-ASR~\cite{GLM-ASR} to generate transcriptions for each 5s training audio clip. To avoid background leakage, for each video segment, we randomly sample a frame containing a face and then crop it as the reference image.

\section{Experiments}
\label{sec:exp}

\subsection{Experimental setup} 

\noindent{\textbf{Implementation Details.}} 
We use OVI 1.0~\cite{low2025ovi} as the base audio-video generation model, which generates 5-second videos at 24 FPS. We crop each reference image to $512\times512$. We utilize 8 H100 GPUs, a batch size of 1 per GPU, and a learning rate of $1e\mbox{-5}$, training over 200,000 steps. To enable large-scale model training and improve computational efficiency, we adopt DeepSpeed~\cite{Rasley2020DeepSpeedSO} as the distributed training framework. The AdamW optimizer is applied, where parameters are set as ${\beta}_1 = 0.9$, ${\beta}_2 = 0.95$, and $\epsilon=1e\mbox{-8}$. 
Besides, the 16kHz audio encoder~\cite{cheng2025mmaudio} is adopted as default. The guidance scales for audio and video are 3.0 and 4.0, respectively. We incorporate LoRA with a rank of 128 for training. Weight coefficients in Eq.~(\ref{eqn:all_loss}) are set as: $\lambda_{V}=1$, $\lambda_{A}=1$, $\lambda_{I^r}=0.1$, and $\lambda_{A^r}=0.1$. During inference, flow-matching sampling is applied with 50 steps.

\noindent{\textbf{Benchmark.}}
Given the lack of a publicly available dataset for sync audio-video customization, we construct a mini benchmark with 100 examples. Specifically, we first reserve 30 persons who were not included in training data of our OmniCustom-1M dataset and then collect 70 videos from YouTube. From these videos, we extract reference images and audio clips for testing. We set the gender ratio of the benchmark to 1:1.

\noindent{\textbf{Evaluation Metrics.}} (i) \textit{FaceSim-Arc} and \textit{FaceSim-Cur}. We extract facial embeddings from the reference image and each generated frame using ArcFace~\cite{deng2019arcface} and CurricularFace~\cite{Huang2020CurricularFaceAC}, respectively. Then, the average cosine similarity between them is computed. 
(ii) \textit{FID}. To evaluate video quality, we employ \textit{FID}~\cite{heusel2017gans}, which computes the feature distribution discrepancy~\cite{szegedy2016rethinking} between face regions in the generated videos and those in reference images.
(iii) \textit{FVD}. We also adopt the widely used \textit{FVD}~\cite{fvd} to measure video quality.
(iv) \textit{CLIP-Text}. We compute the average cosine similarity between the generated frames and the text prompt with CLIP-B~\cite{radford2021learning} image and text embeddings. (v) \textit{Speaker-Sim}. We extract speaker embeddings using WavLM-TDNN~\cite{chen2022wavlm} and compute the cosine similarity between the synthesized and ground-truth speech segments. (vi) \textit{FAD}. We use \textit{FAD}~\cite{fad} to evaluate audio quality. (vii) \textit{Word Error Rate (WER)}. We apply Whisper~\cite{radford2023robust} for automatic speech recognition to obtain text transcriptions of generated audios, and then \textit{WER} can be computed.

% (iv) Temporal Consistency: Following VBench~\cite{huang2024vbench}, we use the CLIP-B model~\cite{radford2021learning} to compute the similarity between each frame and its adjacent frames, as well as the first frame.

\begin{figure*}[t]
  \centering
  \includegraphics[width=1\linewidth]{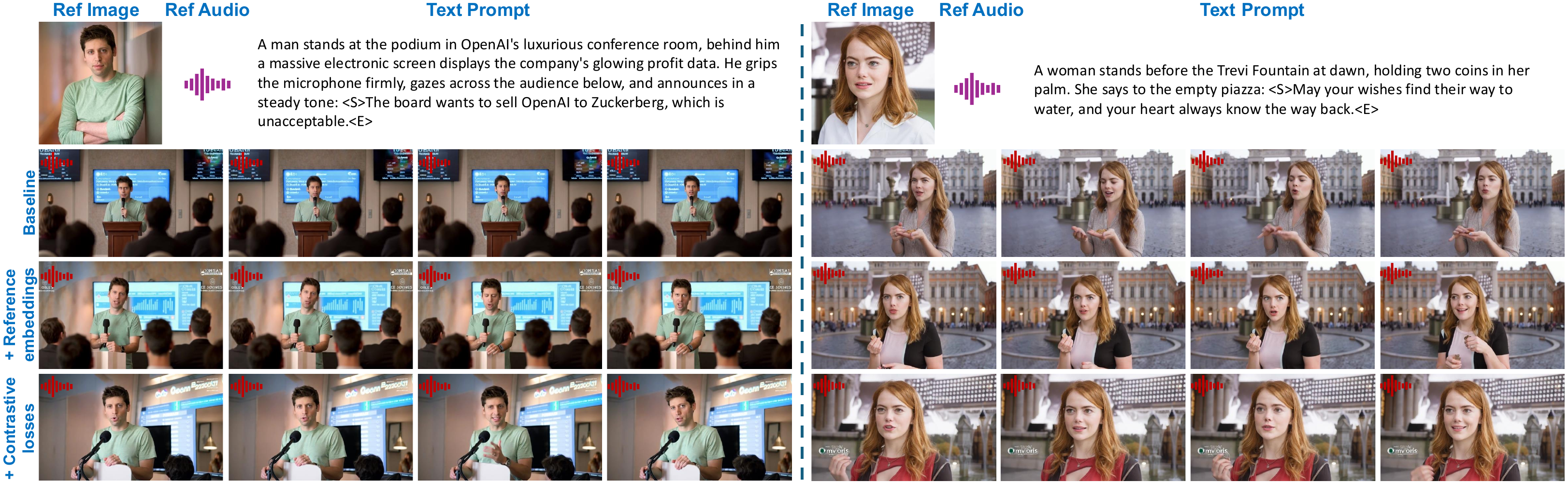}
\vspace{-1em}
  \caption{Ablation study.  
  Face embeddings and the contrastive identity objective boost identity consistency, respectively.
  See the video results for audio details.}
  \label{fig:ablation}
     % \vspace{-0.5em}
\end{figure*}

\subsection{Qualitative Comparison}
We compare our method against typical video customization methods: D-Animator~\cite{he2024id}, ConsisID~\cite{yuan2025identity}, Phantom~\cite{liu2025phantom}, and VACE~\cite{jiang2025vace}. We also include results from audio-driven customization methods: HunyuanCustom~\cite{hu2025hunyuancustom} and Humo~\cite{chen2025humo}. For a fair comparison, we generate the input audio for the audio-driven methods using the spoken content from our benchmark and TTS model CosyVoice~\cite{du2024cosyvoice}. Notably, our method utilizes $<$S$>$ and $<$E$>$ tags to mark spoken content in the prompt, which is removed for the competing methods.

% Specifically, we generate the driven audio using the spoken content from our benchmark and a TTS model CosyVoice~\cite{du2024cosyvoice}, which is then taken as input to these methods~\cite{hu2025hunyuancustom,chen2025humo}. Our method uses <S> and <E> to mark the spoken content in the text prompt. For other methods, we remove this spoken text.

As seen in Fig.~\ref{fig:vis}, our method generates the most visually appealing and identity-preserving video sequence. Specifically, the synthesized faces maintain high fidelity to the reference image throughout the entire sequence. We present more examples in the Appendix. Please refer to our 
supplementary video for timbre imitation ability.

\begin{table*}[t]
% \small
\footnotesize
\centering
\setlength\tabcolsep{0.5pt}
\begin{tabular}{l|l|c|c|c|c|c|c|c|c|c}
\toprule  
\multirow{2}*{Settings}  &\multirow{2}*{Model} 
&\multicolumn{5}{c|}{Video Metrics} &\multicolumn{3}{c|}{Audio Metrics} &\multirow{2}*{\makecell[c]{Background\\Sound}}
\\ 
&     
&\footnotesize{\textit{FaceSim-Arc} $\uparrow$}  
&\footnotesize{\textit{FaceSim-Cur} $\uparrow$} 
&\footnotesize{\textit{FID} $\downarrow$} 
&\footnotesize{\textit{FVD} $\downarrow$} 
&\footnotesize{\textit{CLIP-Text} $\uparrow$} 
&\footnotesize{\textit{Speaker-Sim} $\uparrow$}
&\footnotesize{\textit{FAD} $\downarrow$} 
&\footnotesize{\textit{WER} (\%) $\downarrow$}  
% & \textit{Background Sound}
\\
\hline
\multirow{4}{*}{\makecell[l]{Typical \\Video\\Customization}}
 &ID-Animator~\cite{he2024id}    &0.33  &0.35 & 134.09  &637.93 &25.40  &- &-  &- &-\\
 &ConsisID~\cite{yuan2025identity}   &0.49 &0.50  &174.13  &\underline{471.46}  & 27.05   & - &- &-  &-\\
 &Phantom~\cite{liu2025phantom}  & \underline{0.56}  & \underline{0.57} &169.69  &478.36  &26.72  &- &-  &-  &-\\
 &VACE~\cite{jiang2025vace}  & 0.22  &0.23   & 183.85 &575.09  &\underline{28.03}  &- &- &- &-\\ %\underline{0.935}
\hline
\multirow{3}{*}{{TTS}}
 &F5-TTS~\cite{chen2025f5}   &-  &- &-  &-  &- &\underline{0.55} &3.57 & 3.39  &\ding{55}  \\
 &CosyVoice~\cite{du2024cosyvoice}  &-  & - &-   &-   &- &0.51 &5.20  & 4.77 &\ding{55}\\
 &Fish-speech~\cite{liao2024fish}  &-  &-   & - & -  &- &\textbf{0.60} &\textbf{2.12}  & \textbf{2.26}  &\ding{55}\\
 \hline
\multirow{2}{*}{\makecell[l]{Audio-driven\\Customization}}
&HunyuanCustom~\cite{hu2025hunyuancustom}  &0.53  &0.55  & 130.29  &517.12 &23.98  &-  & - & - &\ding{55} \\
&Humo~\cite{chen2025humo} & 0.50 & 0.52 &181.01  &475.94 &27.23  &-  &-  & - &\ding{55} \\
 \hline
 \rowcolor{mygray} &Ours (\footnotesize{Baseline})   &0.39    &0.41 & 137.18  &559.96 &\textbf{28.31}  &0.29 &4.32 &\underline{2.40}  &\ding{51} \\
  \rowcolor{mygray}   &Ours (\footnotesize{+ Face\&Timbre embedding})  & 0.48   &0.49 & \underline{105.62}  &506.33  &27.68  &0.38 &3.67 &{2.78}  &\ding{51}\\
  \rowcolor{mygray} \multirow{-3}{*}{\makecell[l]{Sync \\Audio-video\\ Customization}}  &Ours (\footnotesize{+ Contrastive learning losses}) &\textbf{0.60}    &\textbf{0.62} & \textbf{95.57}  &\textbf{440.49}  &27.45  &0.47 &\underline{3.44} &2.51  &\ding{51}\\
\bottomrule
\end{tabular}
\caption{Quantitative comparison with existing video customization and TTS methods. The metrics are divided into video and audio parts. The best and second-best results are marked in bold and underlined. In addition, we show whether each method can generate background sound effects.
}
\label{tab:quanti}
\vspace{-1em}
\end{table*}
% Sync Audio-video Customization &Ours   &    & & 142.75 &29.33  & &3.03  &\ding{51}\\
\subsection{Quantitative Comparison}
We generate videos for each identity using 3 random seeds,  bringing a total of 300 videos. The quantitative comparison is presented in Tab.~\ref{tab:quanti}, where our standard OmniCustom is in the \textbf{last row}. We achieve the best $\textit{FID}$ and $\textit{FVD}$, which highlights our advantage in video quality. Further, we outperform all counterparts in \textit{FaceSim-Arc} and \textit{FaceSim-Cur} scores, which shows our superiority on identity preservation. Regarding prompt following, our standard OmniCustom achieves comparable \textit{CLIP-Text} to VACE~\cite{jiang2025vace}, which, however, exhibits the worst identity preservation and video quality.

% To evaluate our timbre cloning capability, we compare with state-of-the-art TTS methods~\cite{chen2025f5,du2024cosyvoice,liao2024fish}. Unlike these vision-free TTS methods, which rely on hundreds of thousands of hours of audio data for timbre cloning, our OmniCustom-1M dataset consists of 2,500 hours of audio-visual data for both identity and timbre customization. Despite this significant difference, we still obtain a competitive timbre cloning result. We believe that a stronger base audio-video model can also enhance our timbre customization performance.

To evaluate our timbre cloning performance, we compare with state-of-the-art TTS methods~\cite{chen2025f5,du2024cosyvoice,liao2024fish} using \textit{Speaker-Sim}. Unlike TTS models, which rely on hundreds of thousands of hours of audio data for timbre cloning, our OmniCustom only applies 2,500 hours of audio-visual data for both identity and timbre customization. Despite this, we still achieve competitive timbre cloning performance with CosyVoice~\cite{du2024cosyvoice}.

Compared with TTS methods, we achieve comparable audio quality \textit{FAD}.
Furthermore, thanks to the capabilities of the base model OVI~\cite{low2025ovi}, we achieve a word error rate \textit{WER} comparable to that of TTS models. Additionally, our method can generate background sounds related to text prompts (\eg, ocean waves) and background music similar to that in the reference audio, while TTS methods cannot. Please refer to our supplementary video for details. We also compare with audio-driven video customization methods~\cite{hu2025hunyuancustom,chen2025humo}, where our OmniCustom outperforms them on all video metrics. Note that these audio-driven methods cannot generate background sounds either. Moreover, to demonstrate the advantage of our sync customization over audio-driven customization, we provide a quantitative comparison on lip synchronization in Appendix.

% Besides, thanks to the capabilities of the base audio-video generation model OVI~\cite{low2025ovi}, we achieve a word error rate \textit{WER} comparable to that of TTS models. Moreover, our method can generate corresponding background sounds, \eg, ocean waves and background music, while TTS methods cannot. Please refer to our supplementary video for details. We also compare with audio-driven video customization methods~\cite{hu2025hunyuancustom,chen2025humo}, where our OmniCustom outperforms them on all video metrics. Note that these audio-driven methods cannot generate background sounds either.

% \vspace{-0.1em}
\subsection{Ablation Study}
We conduct ablation study to validate the effectiveness of face and timbre embeddings and contrastive learning objectives. 
(i) \textbf{Baseline.} We only utilize flow matching losses $\mathcal{L}_{FM}^{V}$ and $\mathcal{L}_{FM}^{A}$, and do not employ face and timbre embeddings. (ii) \textbf{"+ Reference embeddings".} We only use flow matching losses $\mathcal{L}_{FM}^{V}$ and $\mathcal{L}_{FM}^{A}$, where face and timbre embeddings are employed. (iii) \textbf{"+ Contrastive losses".} Compared to (ii), we further apply contrastive learning objectives $\mathcal{L}_{CL}^{I}$ and $\mathcal{L}_{CL}^{A}$ in Eq.~(\ref{eqn:all_loss}) as our standard version.

We provide quantitative ablation results in the last three rows of Tab.~\ref{tab:quanti}. Face embeddings and contrastive identity objective $\mathcal{L}_{CL}^{I}$ can significantly enhance identity preservation and video quality. Furthermore, timbre embeddings and contrastive timbre loss $\mathcal{L}_{CL}^{A}$ boost timbre similarity by 31.0\% and 23.7\%, respectively, where audio quality can also be consistently improved.
Across all three settings, the prompt following metric \textit{CLIP-Text} and word error rate \textit{WER} show little variation. We also show visual ablation results in Fig.~\ref{fig:ablation}, which supports our findings in Tab.~\ref{tab:quanti}. \textbf{Baseline} struggles to maintain the identity information and is prone to generating artifacts in facial details. \textbf{"+ Reference embeddings"} improves the baseline to a great extent while our standard OmniCustom (\textbf{"+ Contrastive losses"}) achieves the best generation.

\subsection{User Study}
We conduct a user study with 20 participants, where each participant evaluated 10 randomly sampled videos from our benchmark. The study adopts a standard two-alternative forced-choice paradigm. Specifically, each participant is provided with a reference image and two customized outputs, where one is generated by our OmniCustom while the other by the competing method. Participants are required to select the superior output with respect to \textit{ID Consistency}, \textit{Audio-video Sync}, and \textit{Video Quality}. The selection ratios are summarized in Tab.~\ref{tab:user}, where our OmniCustom method achieves a win rate exceeding 50\% (random chance) for all competing approaches. This demonstrates the superiority of our method in all three key metrics. In particular, we achieve better audio-visual synchronization than audio-driven customization methods~\cite{hu2025hunyuancustom,chen2025humo}.

\begin{table}[t]
\small
\centering
\setlength\tabcolsep{0.5pt}
\begin{tabular}{l|c|c|c}
\toprule  
{Ours \textit{vs.}} 
&\makecell[c]{\textit{Identity}\\\textit{Consistency}}  $\uparrow$ 
% &\textit{ID Consistency} $\uparrow$ 
&\makecell[c]{\textit{Audio-video} \\\textit{Sync}}  $\uparrow$  
% &\textit{Audio-video Sync} $\uparrow$ 
% &\textit{Video Quality} $\uparrow$ \\
&\makecell[c]{\textit{Video} \\ \textit{Quality}} $\uparrow$ \\
\hline
ID-Animator~\cite{he2024id}       &95\%   &-  &96\%\\
ConsisID~\cite{yuan2025identity}  &91\%  &-  &94\%\\
Phantom~\cite{liu2025phantom}     &74\%   &-   &86\%\\
VACE~\cite{jiang2025vace}         &90\%  &-  &91\%\\
\hline
HunyuanCustom~\cite{hu2025hunyuancustom} & 81\% &88\%  &85\%\\
Humo~\cite{chen2025humo}&86\%  & 79\% &83\%\\
% \hline
% Ours  & &  &\\
\bottomrule
\end{tabular}
  % \vspace{-0.8em}
\caption{User Study. We achieve a higher preference rate.
}
\label{tab:user}
  % \vspace{-2.2em}
\end{table}

%~\cite{anyv2v}
%~\cite{i2vedit}
%~\cite{videoshop}

 % rating them on a 5-point scale (1-5) across three dimensions: identity consistency, audio-video synchronization, and video quality. The final score for each dimension was computed by averaging ratings across all participants and videos. As shown in Tab.~\ref{tab:user}, our method outperforms existing approaches, earning higher scores across all evaluated aspects.

\section{Conclusion}
We propose a new task of \textit{sync audio-video customization}, which 
aims to synchronously generate a video that preserves the visual identity from the reference image $I^{r}$ and an audio track that mimics the timbre of the reference audio $A^{r}$, while allowing the speech content to be freely given by a textual prompt. Building upon a state-of-the-art sync audio-video generation model, we propose OmniCustom, a novel framework which embeds identity and audio information into the original video and audio branches through self-attention layers, respectively, where two independent LoRAs are integrated into the QKV projections of reference tokens. To further boost the fidelity of identity preservation and timbre imitation, we design two complementary contrastive learning objectives, which maximize the dissimilarity between predicted flows under reference-guided and non-reference conditions. Extensive experiments demonstrate that our OmniCustom achieves state-of-the-art performance in identity-preserving text-to-video generation, while simultaneously realizing high-fidelity timbre cloning. 

% While our method achieves promising results, it still faces some limitations. For example, due to the constraints of the base joint audio-video generation model, our OmniCustom currently supports only English and video generation of 5-second duration.

% \newpage

{
    \small
    \bibliographystyle{ieeenat_fullname}
    \bibliography{main}
}

\clearpage
\clearpage
\setcounter{page}{1}
\setcounter{figure}{0}
\setcounter{table}{0}
\renewcommand{\figurename}{\textbf{A-Fig.}}
\renewcommand{\tablename}{\textbf{A-Tab.}}

\appendix

\twocolumn[{%
\renewcommand\twocolumn[1][]{#1}%
\maketitlesupplementary
\begin{center}
    \centering
    \captionsetup{type=figure}
    \includegraphics[width=\textwidth]{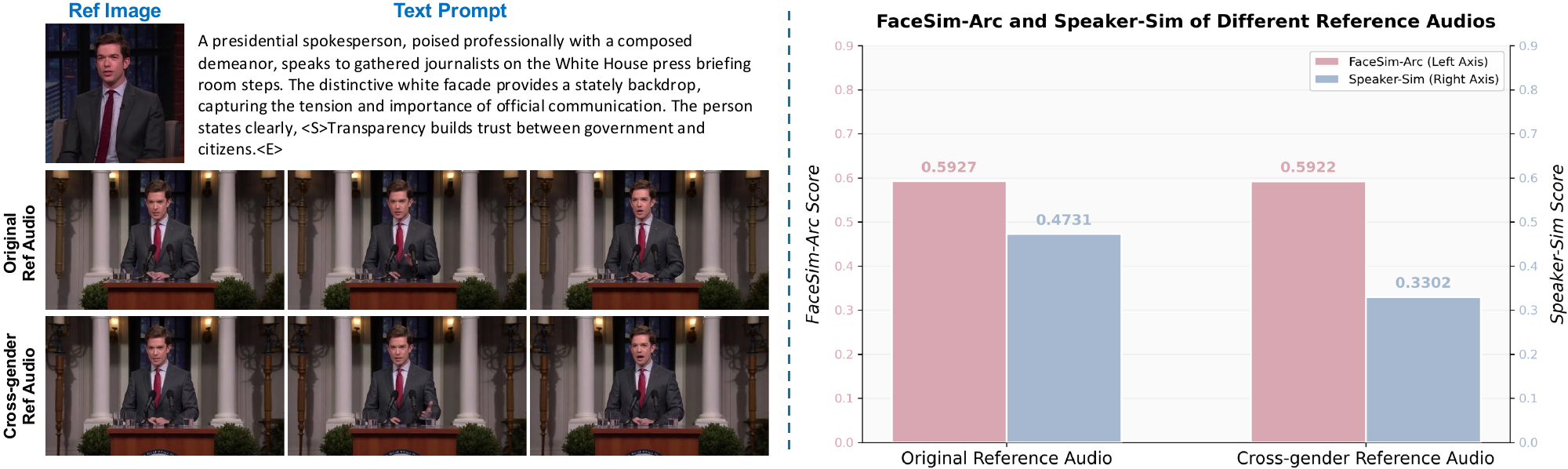}
       \caption{Our OmniCustom can perform cross-gender audio customization.
       }  
    \label{fig:cross}
\end{center}%
}]

\maketitle

This Appendix is organized as follows. Appendix~\ref{sec:A} elaborates on the construction details of our dataset and benchmark. Appendix~\ref{sec:B} discusses further applications of our method. Appendix~\ref{sec:C} and Appendix~\ref{sec:D} provide more quantitative comparisons and visualization results.

% Finally, Appendix~\ref{sec:E} discusses the broader impacts of this work.

% \begin{teaserfigure}
%   \includegraphics[width=0.9\textwidth]{../figs/data_distribution.pdf}
%        \centering
%   \caption{Statistics of our proposed dataset OmniCustom-1M, in terms of age and gender.
% }
%   \label{fig:data}
%   \vspace{2em}
% \end{teaserfigure}

\begin{figure*}[htbp]
  \centering
  \includegraphics[width=0.8\linewidth]{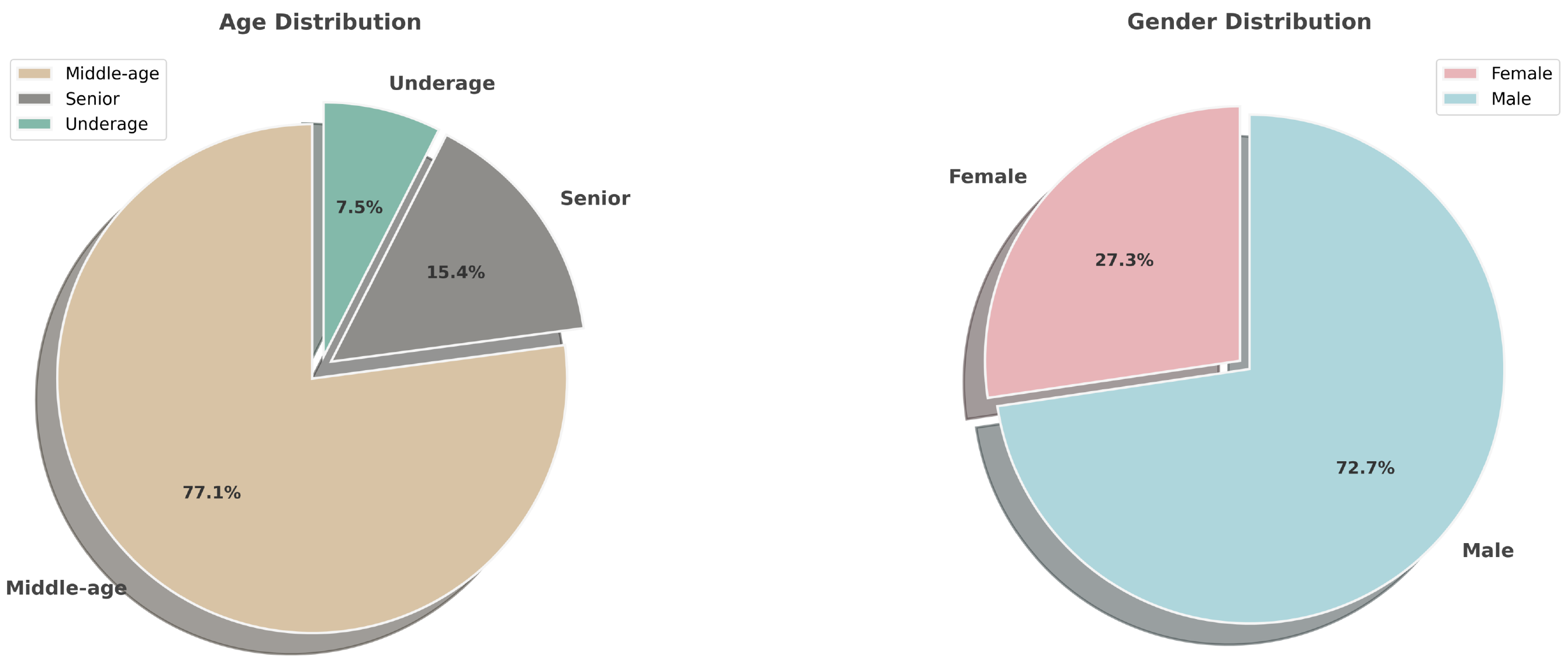}
    \vspace{-0.6em}
  \caption{Statistics of our proposed dataset OmniCustom-1M, in terms of age and gender.
}
\label{fig:data}
  \vspace{1em}
\end{figure*}
\begin{figure*}[htbp]
  \centering
  \includegraphics[width=0.9\linewidth]{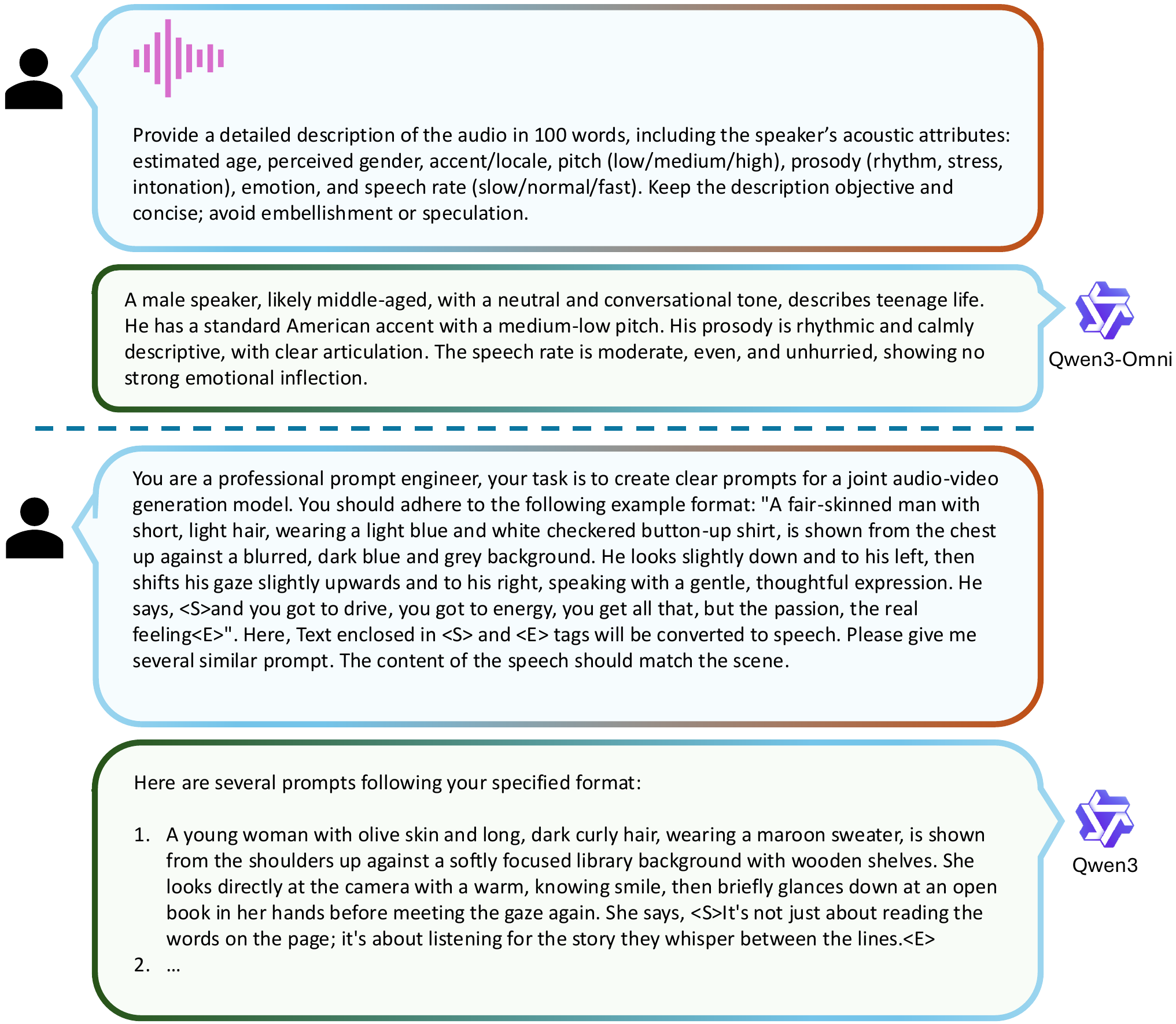}
        \vspace{-1em}
  \caption{Prompt used for audio caption and benchmark construction.
}
\label{fig:caption}
\end{figure*}

  % \vspace{-1.0em}
\begin{figure*}[hbtp]
  \centering
\includegraphics[width=1\linewidth]{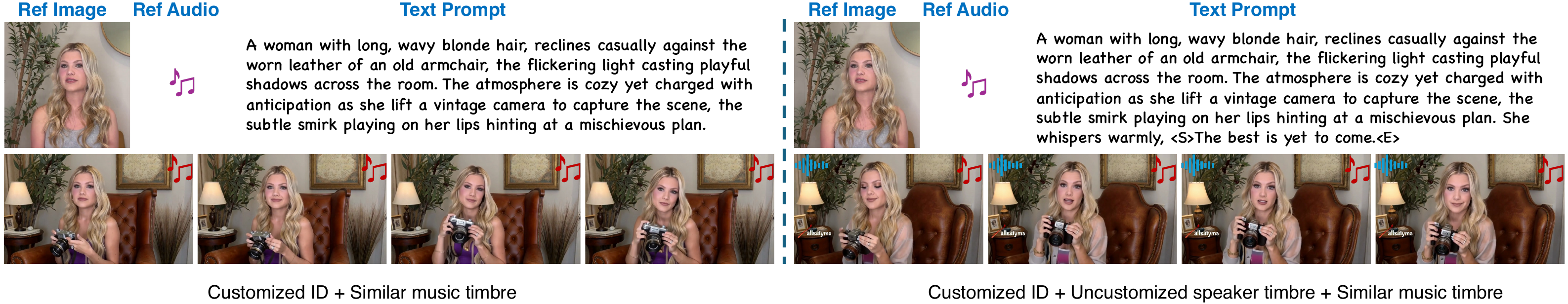}
  \caption{When reference audio is a pure music clip, our OmniCustom can generate similar background music.
}
\label{fig:music}
% \vspace{-1em}
\end{figure*}

\section{More details of Dataset and Benchmark}
\label{sec:A}
\subsection{Audio Captioning and Textual Prompts}
We elaborate on the prompts for audio captioning in our OmniCustom-1M dataset, as well as those designed for benchmark construction. As shown in the top half of A-Fig.~\ref{fig:caption}, we utilize the 30B version of Qwen3-Omni~\cite{Qwen3-Omni} to generate a set of clip-level audio captions. These annotations provide a structured, multifaceted description of each audio clip, facilitating detailed audio analysis and control. Furthermore, as shown in the bottom half of A-Fig.~\ref{fig:caption}, we leverage Qwen3~\cite{yang2025qwen3} to generate textual prompts for our benchmark.

\subsection{Statistics of Our Dataset}
A-Fig.~\ref{fig:data} illustrates the gender and age distributions in OmniCustom-1M. We categorize ages into three groups: young, middle-aged, and senior. Although not perfectly uniform, the dataset is sufficiently diverse to facilitate the novel task of \textit{sync audio-video customization}.

\section{Additional applications}
\label{sec:B}
In A-Fig.~\ref{fig:music}, we provide the results of using a pure music clip as reference audio. OmniCustom can generate similar background music in personalized results, which is not supported by existing audio-driven customization methods. When the text prompt does not contain spoken content, our customized results merely yield similar music timbre. In contrast, when the text prompt contains spoken content, the corresponding result show uncustomized speaker timbre and similar music timbre. Please see our supplementary video for audio details.

\begin{table}[t]
\setlength\tabcolsep{12pt}
    \centering
    \small
    \begin{tabular}{l|c|c}
    \toprule  
      Settings   &\textit{LSE-C} $\uparrow$  & \textit{LSE-D} $\downarrow$ \\
    \hline
HunyuanCustom~\cite{hu2025hunyuancustom}   &4.87  &10.47  \\
Humo~\cite{chen2025humo}  &\underline{5.42}  &\underline{9.31}  \\
\hline
 \rowcolor{mygray}
Ours &\textbf{5.76} &\textbf{8.53}  \\
     \bottomrule
    \end{tabular}    
    \caption{Quantitative comparison with existing audio-driven video customization methods on lip synchronization.}
    \label{tab:sync}
\end{table}

% \begin{table}[t]
% \small
% \centering
% % \setlength\tabcolsep{3pt}
% \begin{tabular}{l|c|c}
% \toprule  
% {Settings} 
% &LSE-C $\uparrow$   
% &LSE-D $\downarrow$ \\
% \hline
% HunyuanCustom~\cite{hu2025hunyuancustom} & 81\% &88\% \\
% Humo~\cite{chen2025humo}&86\%  & 79\% \\
% \hline
% \rowcolor{mygray} Ours  & & \\
% \bottomrule
% \end{tabular}
% \caption{Sync.
% }
% \label{tab:sync}
% \end{table}
% \input{table_tex/user_study}

\begin{figure*}[t]
  \centering
  \includegraphics[width=1\linewidth]{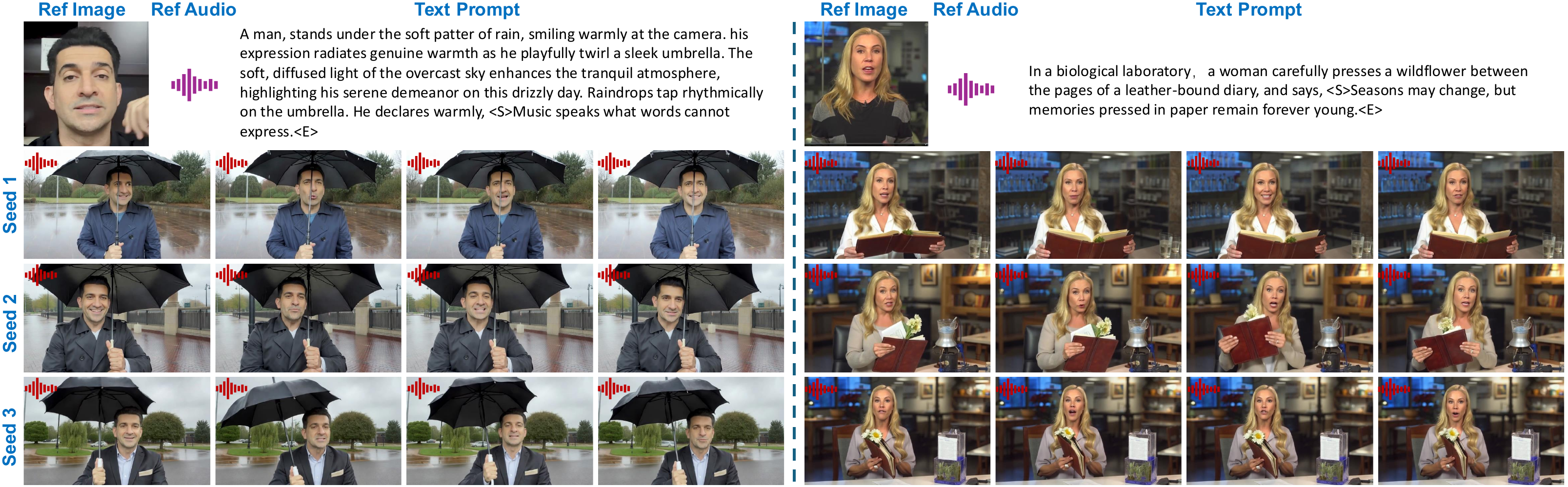}
        % \vspace{-2em}
  \caption{Our results using three different random seeds. Our method demonstrates the diversity of background and clothing.
}
\label{fig:multiseed}
\end{figure*}

Our method can also perform cross-gender reference audio customization. As illustrated in A-Fig.~\ref{fig:cross}, pairing a male reference image with a female reference audio leads to a slightly lower \textit{Speaker-Sim} score compared to the original female reference audio. 
We attribute such degradation to the model implicitly encoding some priors regarding the gender of reference identity and its corresponding timbre range. Notably, the change in reference audio does not compromise the identity preservation.

\begin{figure}[h]
  \centering
  \includegraphics[width=1\linewidth]{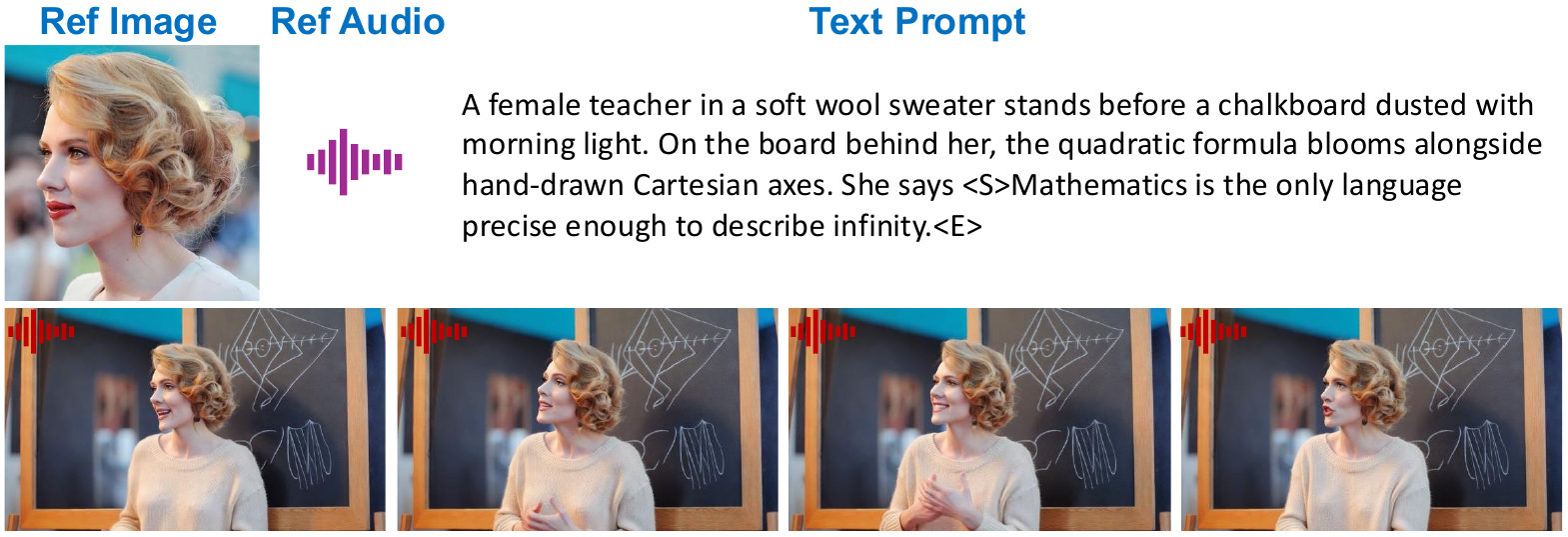}
        % \vspace{-2em}
  \caption{Failure case of using a profile face as the reference image. Our OmniCustom 
  struggles to preserve the identity of profile faces.
}
\label{fig:fail}
\end{figure}

\section{More Quantitative Results}
\label{sec:C}

Recall that in Tab.~1 of our main paper, we present video and audio metrics regarding identity preservation, video/audio quality, and timbre cloning. Here, to quantitatively evaluate lip synchronization and mouth shape, we use the distance score (\textit{LSE-D}) and confidence score (\textit{LSE-C}) to assess the perceptual differences in the mouth shapes generated by Wav2Lip~\cite{prajwal2020lip}. The results are reported in A-Tab.~\ref{tab:sync}. Compared with current audio-driven customization methods ~\cite{hu2025hunyuancustom,chen2025humo}, our OmniCustom achieves the highest confidence score and the lowest distance score, which demonstrates the superiority of our method in lip synchronization.

% We argue that audio-driven customization methods learn unidirectional conditional distribution $p(V|A)$, while our sync customization method models the joint distribution $p(V,A)$, which leads to coherent audiovisual performance due to richer information.

% \subsection{User Study}
% We conduct a user study with 20 participants, where each participant evaluated 10 randomly sampled videos from our benchmark. The study adopts a standard two-alternative forced-choice paradigm. Specifically, each participant is provided with a reference image and two customized outputs, where one is generated by our OmniCustom while the other by the competing method. Participants are required to select the superior output with respect to \textit{ID Consistency}, \textit{Audio-video Sync}, and \textit{Video Quality}. The selection ratios are summarized in A-Tab.~\ref{tab:user}, where our OmniCustom method achieves a win rate exceeding 50\% (random chance) for all competing approaches. This demonstrates the superiority of our method in all three key metrics. In particular, we achieve better audio-visual synchronization than audio-driven customization methods~\cite{hu2025hunyuancustom,chen2025humo}.

\begin{figure*}[t]
  \centering
  \includegraphics[width=1\linewidth]{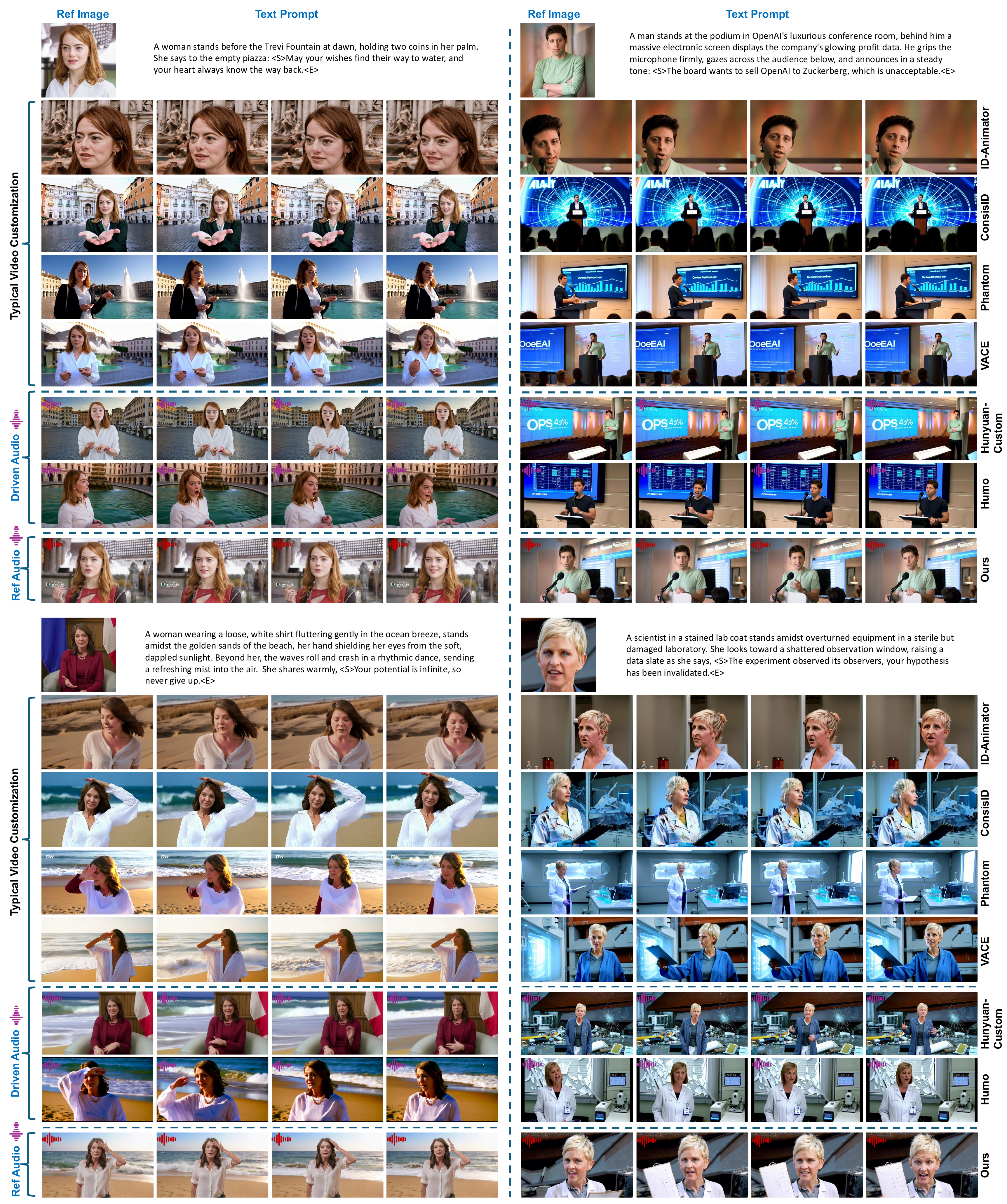}
        \vspace{-2em}
  \caption{More qualitative comparison with existing customization methods.
}
\label{fig:vis1}
\end{figure*}
\begin{figure*}[t]
  \centering
  \includegraphics[width=1\linewidth]{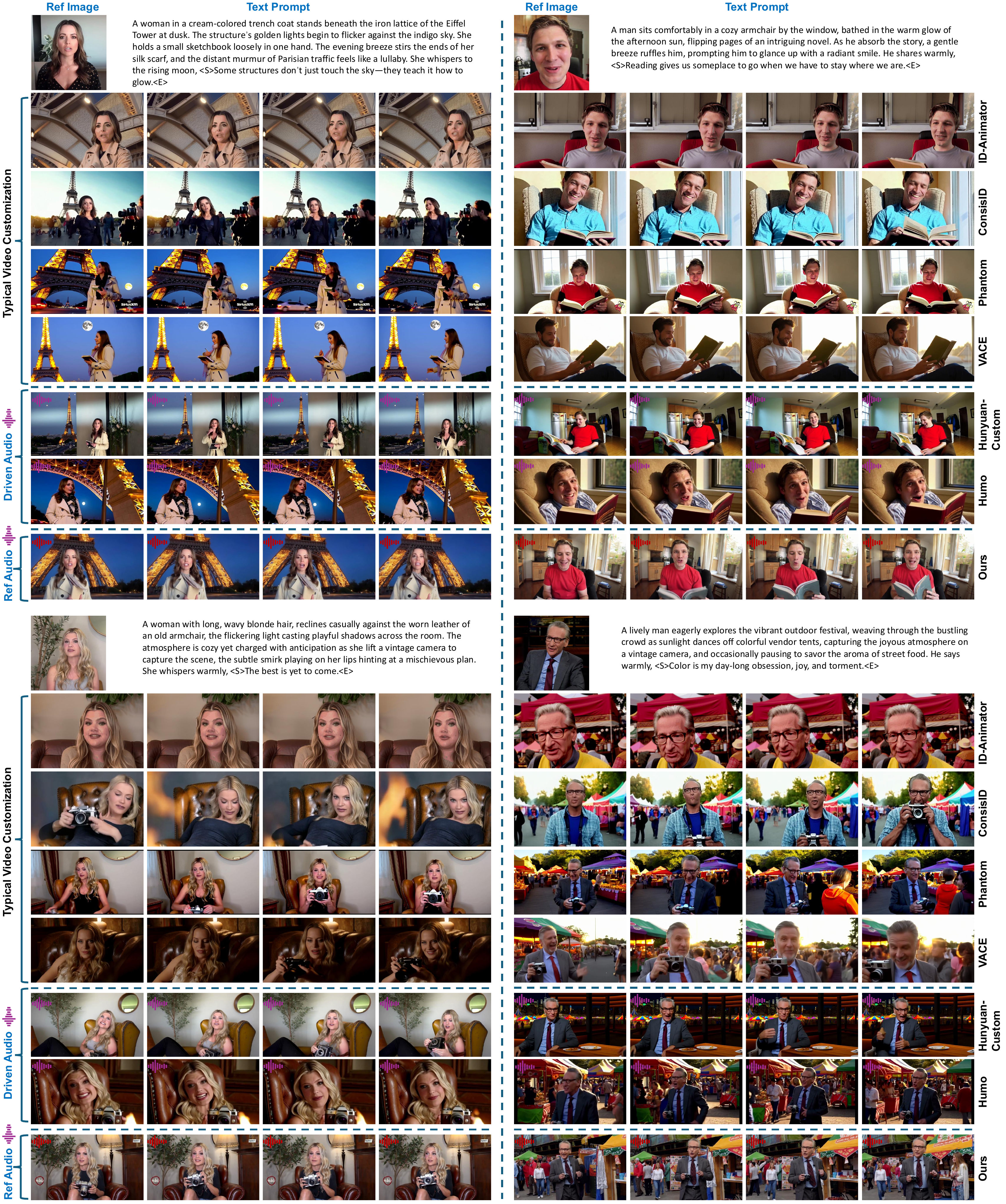}
        \vspace{-2em}
  \caption{More qualitative comparison with existing customization methods.
}
\label{fig:vis2}
\end{figure*}

\section{More Qualitative Results}
\label{sec:D}

\subsection{Multi-seed Results}
In A-Fig.~\ref{fig:multiseed}, we display the results when using the same reference image and audio and text prompt with three different random seeds. While maintaining facial fidelity, our method exhibits outstanding diversity in background and clothing.

\subsection{Limitations}
While our method achieves promising results, it still faces some limitations. For example, due to the constraints of the base joint audio-video generation model, our OmniCustom currently supports only English and video generation of 5-second duration.

In addition, we present the results of using a profile face as the reference image in A-Fig.~\ref{fig:fail}. It can be observed that our method tends to preserve the pose for a given profile face, which leads to relatively weak identity preservation. We believe such limitation can be addressed by employing multi-view reference images during training, which we leave as future work.

\subsection{More Comparisons With Existing Methods}
In A-Fig.~\ref{fig:vis1} and A-Fig.~\ref{fig:vis2}, we present additional qualitative comparisons with typical video customization methods~\cite{he2024id,yuan2025identity,liu2025phantom,jiang2025vace} and audio-driven customization methods~\cite{hu2025hunyuancustom,chen2025humo}. Our method demonstrates superior identity preservation and prompt following capabilities. Meanwhile, our results show the best video quality.

\end{document}